\documentstyle[twocolumn,prl,aps,epsfig]{revtex}

\begin{document}
\draft
\title{Diffusion-Reorganized Aggregates: Attractors in Diffusion Processes?}
\author{Marcel Filoche and Bernard Sapoval}
\address{Laboratoire de Physique de la Mati\`{e}re Condens\'{e}e\\
Ecole Polytechnique, CNRS\\
91128 Palaiseau C\'{e}dex, France.}
\date{\today }
\maketitle

\begin{abstract}
A process based on particle evaporation, diffusion and redeposition is
applied iteratively to a two-dimensional object of arbitrary
shape. The evolution spontaneously transforms the object morphology,
converging to branched structures. Independently of initial geometry,
the structures found after long time present fractal geometry with a
fractal dimension around 1.75. The final morphology, which constantly
evolves in time, can be considered as the dynamic attractor of this
evaporation-diffusion-redeposition operator. The ensemble of these
fractal shapes can be considered to be the {\em dynamical
equilibrium} geometry of a diffusion controlled self-transformation
process.
\end{abstract}


% 05.40.+j	Fluctuation phenomena, random processes,
%		and Brownian motion
% 61.43.Hv	Fractals; macroscopic aggregates
%		(including diffusion-limited aggregates)
% 64.60.Cn	Order-disorder transformations; statistical
%		mechanics of model systems

This letter reports the discovery of a diffusion mediated process
which spontaneously builds a dynamic fractal {\em equilibrium}
structure, in contrast with fractal morphologies linked to {\em
far-from-equilibrium} processes \cite{Meakin98,Vicsek92,Bak87}. The
process is a surface to surface evaporation-diffusion-condensation
process which conserves the total mass of the system. During the time
evolution, one observes a progressive transformation of the surface
through bulk diffusion. After a long relaxation period, the system
reaches a dynamical fractal structure. This structure appears as the
final equilibrium state towards which {\it any} initial morphology of
$M$ particles will converge after sufficient
evaporation-diffusion-condensation iterations. It can then be
considered as a general statistical attractor for that specific
dynamic process.

The underlying ideas that have suggested this study come from our
knowledge of the basic mathematical objects which govern the exchange
of Laplacian driven currents across irregular (or fractal) interfaces
(as, for example, in the study of irregular or fractal
electrodes). This problem can be mapped onto the study of the transfer
of Brownian particles across irregular membranes with finite
permeability \cite{Sapoval87}. In this process, Brownian particles
strike an irregular surface where they are absorbed with finite
probability. When reflected, the particles undergo successive random
paths, hitting and hitting again the static surface, until they are
finally absorbed. Halsey first indicated that the response of such
systems depends on the probability that a particle starting on the
interface comes back to it \cite{Halsey92}. Generalizing these ideas,
it has been recently shown that the Laplacian transfer across
irregular interfaces is controlled by a single linear operator
$\tilde{Q}$ which maps the {\it static} surface onto itself through
effective ''Brownian bridges'' \cite{Filoche99_1}. In this context,
each surface has an operator $\tilde{Q},$ which is symmetric and
positive.

\begin{figure}[tbp]
\centerline{ \psfig{file=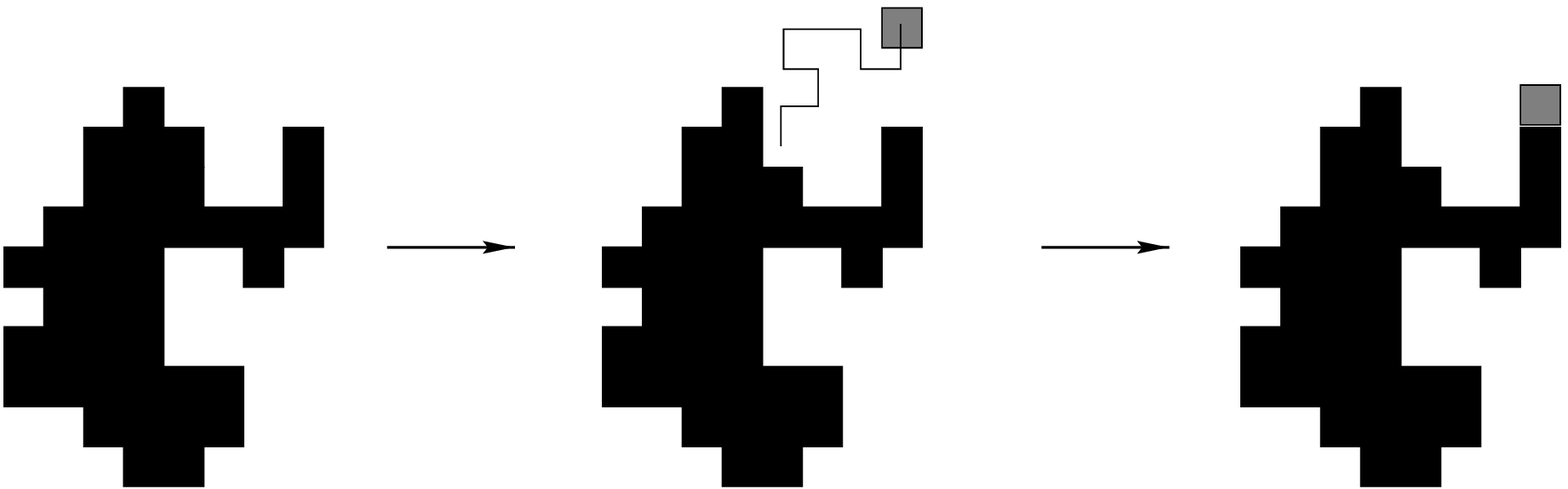,width=.4\textwidth}}
\smallskip
\caption{Schematic of the self-diffusion reorganization
process. Step~1: a particle on the boundary of the structure is
randomly chosen to be dissolved, according to the rule that it does
not break the connectivity of the structure. Step~2: it can go back to
its original position or starts a random walk on a square lattice. If
the walk brings the particle to a surface site, the particle sticks to
the structure under the condition that the new position conserves the
simple connectivity of the structure.}
\label{fig:process}
\end{figure}

The question arises naturally of the link between the Brownian bridges
and a real diffusion process. This led us to study a dynamical process
in which the notion of a $\tilde{Q}$ {\it operator}, which maps the
surface onto itself through average diffusion, is transformed into a
real {\it operation} which transforms the surface through discretized
diffusion \cite{Filoche99_2}. The evolution mechanism then proceeds in
the following steps (Fig.~\ref{fig:process}):

\begin{enumerate}
\item A particle is chosen at random on the surface of an initial
structure and is dissolved, occupying an empty site next to its
initial position.

\item The dissolved particle may jump back to its original site or
start a Brownian random walk on a square lattice which represents the
outer medium. The time step for each jump is $\tau = 1.$ This motion
is stopped when the particle hits the structure again as it sticks on
first hit \cite{Botet85}.

\item The process, which conserves the total mass, is then
iterated. For practical reasons, the system is enclosed in a large
square window: whenever a diffusing particle strikes the window
boundary, this particle is reinitiated on another site of this
external box, following a probability law that simulates the random
walk in the infinite outer medium. As the probability to return to a
starting site is equal to 1 in $d=2$, this does not modify the
morphologies which are generated by the process. The existence of the
external box only perturbs the time scale so that the number of
computer steps is not proportional to a real time. For this reason,
the number of steps will be called pseudo-time $t$.
\end{enumerate}

The dissolution-recondensation process rapidly creates several
branches. If particles of these branches were allowed to dissolve, it
would lead to a progressive splitting of the structure into two, then
many, disconnected parts. In order to avoid this progressive
disconnection, one imposes the structure to remain connected
throughout the process. The only particles allowed to evaporate are
then the particles pertaining to the surface of the structure and that
are not ``red particles'': the red particles are defined as the
particles which, if eliminated, would disconnect the structure in two
or more distincts parts. The particles allowed to evaporate are
therefore called ``blue particles''. We also impose that an evaporated
particle cannot stick on a site where it would disconnect the outer
medium. In fact, one could release this last constraint since it does
not modify the long term evolution of the structure. It simply avoids
to create temporary ``lakes'' in which particles evaporating and
redeposited would greatly reduce the speed of evolution of the
structure. Particles on the internal surface of these lakes would
spend most of the time to go back and forth until the lake opens
spontaneously.

\begin{figure}[tbp]
\centerline{
\psfig{file=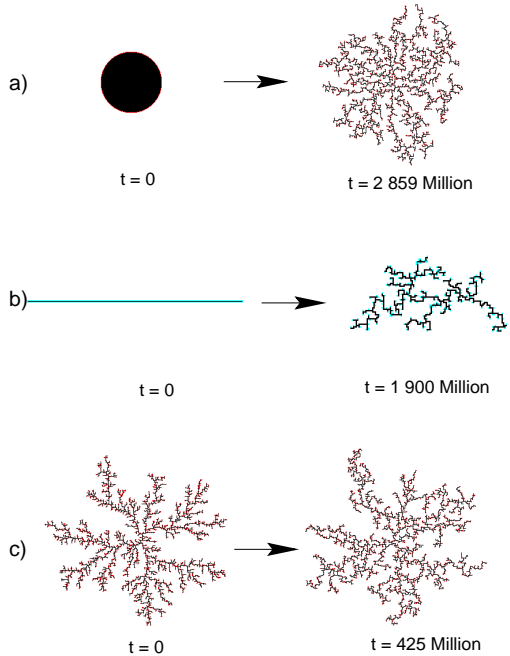,width=.4\textwidth}}
\smallskip
\caption{Three examples of the spontaneous evolution of morphologies
towards statistical equilibrium shapes under the diffusion limited
reorganization process. Case~(a): the initial morphology is dense with
dimensionality $D_f = 2$, and contains 7800 particles. Case~(b):
initial morphology with dimensionality $D_f = 1$, 983
particles. Case~(c): initial DLA morphology with dimensionality $D_f =
1.65$, 8000 particles. The final morphologies look all the same and
keep the same statistical characteristics after being formed. The time
for evolution strongly depends on the initial shape.}
\label{fig:morphevol}
\end{figure}

The results of the spontaneous evolution of structures obtained with
different initial morphologies are shown in
Fig.~\ref{fig:morphevol}. The left column shows the initial shape and
the corresponding final morphologies are shown on the right. In
case~(a), the initial morphology is a compact dense structure
described by a dimensionality equal to~2. In case~(b), the starting
shape is a line with dimensionality~1. Finally, in case~(c), the
initial morphology is an ordinary Diffusion-Limited Aggregate (DLA),
first introduced by Witten and Sander \cite{Witten81}.

In all three cases (a), (b) and (c), the final morphologies present
branched structures, like DLA but less open. The complete statistical
study of the morphologies is difficult because the process is very
time consuming in the computer. For instance, the time necessary to
reach the morphology shown in Fig.~\ref{fig:morphevol}(a) is of the
order of several weeks of CPU time on a Hewlett-Packard C160
workstation. The similarity in the final shapes, independent of
initial geometry, suggests that this class of morphologies is an
attractor for such self-reorganization processes. We call these
structures ``Diffusion-Reorganized Aggregates'' (or ``DRA'').

\begin{figure}[tbp]
\centerline{
\psfig{file=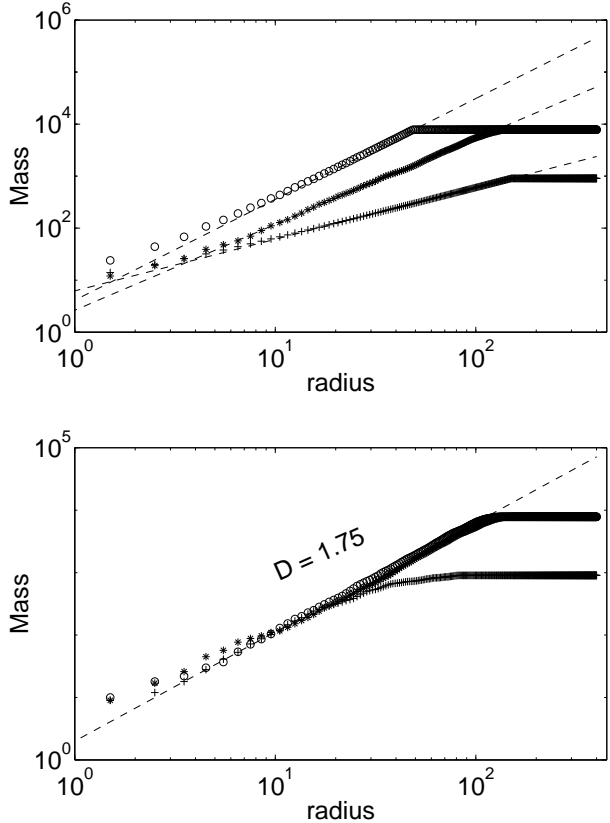,width=.45\textwidth}}
\medskip
\caption{Evolution of the fractality. Top: mass to radius relation for
initial morphologies (a), (b) and (c) (a $\rightarrow$ o, b
$\rightarrow$ +, c $\rightarrow$ *). Bottom: mass to radius relation
for the corresponding stationary states. This plot indicates that,
independent of the initial dimension, the stationary dimensions are
very close.}
\label{fig:Mr}
\end{figure}

The fact that morphologies are ``attracted'' by the fractal final form
is also indicated in Fig.~\ref{fig:Mr}. The figure gives the
determination of the fractal dimension of the final morphologies
starting respectively with initial dimension $D_f = 2$, $D_f = 1$, and
$D_f=1.65$ (Fig.~\ref{fig:morphevol} (a), (b) and (c)). Although the
initial dimension are very different, the final plots are within
reasonable uncertainty concentrated and compatible with $D_f = 1.740
\pm 0.02$, a value which is significantly different from the fractal
dimension of lattice animals \cite{Wessel92} and compatible with that
of the branched structures which are the results of a
far-from-equilibrium process \cite{Lucena94}.

The dynamics of the restructuration process has several specific
properties. To follow the dynamics, one can compute two memory
functions. First, a density-density memory function. For this purpose,
the structure is characterized by a number $s_{i}$ on each site $i$ of
the lattice, with $s_{i}=1$ if the site is occupied by the structure
and $0$ if not. Calling $M$ the mass of the initial structure, which
is equal to the number of occupied sites, this memory function can be
defined as:

\begin{equation}
C(t)=\frac{1}{M}\sum_{{\rm lattice~~sites}~i}s_{i}(0)~s_{i}(t)
\label{eq:corrC}
\end{equation}

\begin{figure}[tbp]
\centerline{ \psfig{file=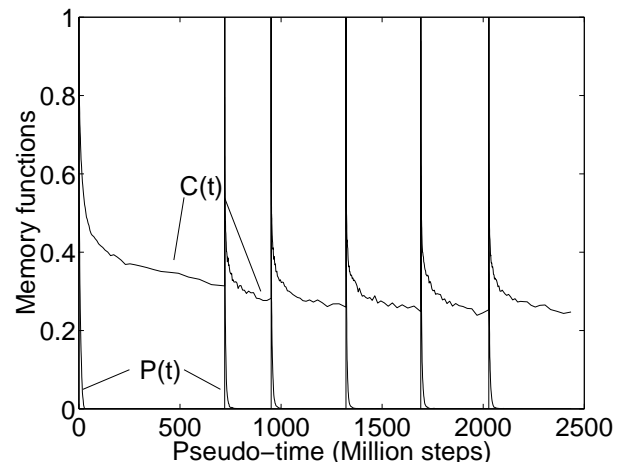,width=.45\textwidth}}
\medskip
\caption{Time evolution of the memory functions $C(t)$ and $P(t)$
during the evolution of the disk towards DRA structure
(Fig~\ref{fig:morphevol}a). These functions are defined by
Eq.~\ref{eq:corrC} and Eq.~\ref{eq:corrP}. Both functions are reset
five times during the evolution (positions of the peaks). After one
million time steps, the correlation $C(t)$ is less than $1/2$,
indicating that more than half of the initial points have moved and no
more belong to the initial structure.}
\label{fig:corr}
\end{figure}

This first function does not discriminate between particles as
$s_{i}(0)~s_{i}(t)$ is equal to 1 if a lattice site $i$ is occupied
both at pseudo-times $0$ and $t$ (by any particle) and 0
otherwise. The pseudo-time evolution of $C(t)$ is shown in
Fig.~\ref{fig:corr}. One observes a memory loss which does not go to
zero. There can be two reasons for incomplete decorrelation. First, it
could happen that a part of the object is not restructured. Second, it
could be that the density-density memory does not vanish due to an
average statistical overlap, even if the structure has completely
forgotten its initial geometry. To clarify this point, one can compute
a particle-particle memory function which characterizes the speed at
which the particles of a given structure finally move from their
initial positions. This second function is

\begin{equation}
P(t)=\frac{1}{M}\sum_{particles~j} \delta(\vec{r_j} (0) - \vec{r_j}
(t)) \label{eq:corrP}
\end{equation}

\noindent
where $\vec{r_j} (t)$ is the position of the particle $j$ at time $t$.\\

The evolution of $P(t)$ is also given in Fig.~\ref{fig:corr}. One
observes that the particle-particle memory is lost to zero very
rapidly. When it has reached very small values, one can say that the
structure has completely lost the memory of its microscopic
configuration. Apart from these two rapid processes, one observes in
Fig.~\ref{fig:corr} the very slow global evolution. Note that the DRA
geometry loses constantly its memory whereas the DLA growth keeps the
total memory of its previous structure.

\begin{figure}[tbp]
\centerline{
\psfig{file=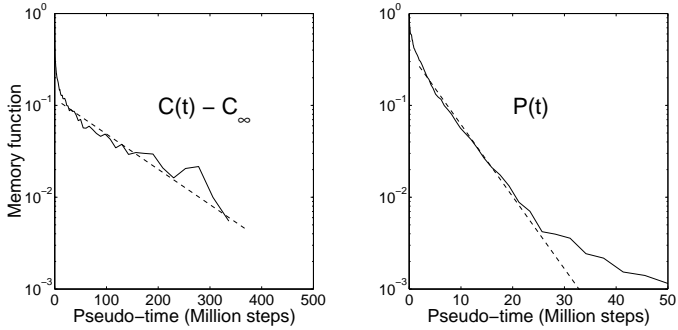,width=.5\textwidth}}
\medskip
\caption{Detailed behavior of the memory functions decays $C(t) -
C(\infty)$ and $P(t)$. In both cases, the final decay is found to be
exponential, with characteristic times of $\tau_C \approx 1.1~10^8$
and $\tau_P \approx 5.10^6$. This suggests a classical equilibrium
situation.}
\label{fig:expcorr}
\end{figure}

A more detailed characterization of the rapid regimes is displayed in
Fig.~\ref{fig:expcorr}, where the decorrelations are found to be
exponential, a non-usual time behavior in this context.

Those results, together with the fractality evolution of
Fig.~\ref{fig:Mr}, indicate that the final DRA morphologies behave as
stable fixed points. The constantly changing DRA can then be viewed as
a statistical equilibrium state. In that sense, it is analogous to the
homogeneous state of an ideal gas, where the microscopic interactions
between particles lead to a statistical macroscopic equilibrium,
although it is constantly changing its microscopical structure.

One should also note that the transformation of the $\tilde{Q}$ {\it
operator}, which maps the surface onto itself through average
diffusion, into a real {\it operation} is analogous to the dielectric
breakdown model of DLA \cite{Pietronero}. In the DBM model, the {\it
operator}, that maps the source at infinity onto the growing
morphology through average Brownian paths, is transformed into a real
stochastic {\it operation}. In the model presented herein, the
source is the object itself.

It is also interesting to return to the exact concept of the
self-transport operator attached to a given morphology
\cite{Filoche99_1}. Extending the notion of diffusive self-transport
in this context, it might be that DRA shapes could be considered as
''{\it eigenshapes}'' of these operators. By this, it is suggested
here that the self-transport operators attached to these structures
transform those structures into themselves, of course in a statistical
sense.

All these results should be confirmed in the future through extensive
numerical simulations and extension to diffusive self-transport in
$d=3$. Some future extensions of this work can be envisaged. First,
one should verify the universal character of this result. Second, if
surface energy was included, it would be interesting to study how the
equilibrium morphology would depend on its value. Last, one should
study whether the self-transport operators $\tilde{Q}$ of these
geometries possess some extremal properties.

In summary, it has been shown that diffusive self-transport
statistically modifies an object morphology in an irreversible
manner. This irreversible process leads to a class of branched
structures, called DRA, with fractal dimension around 1.75. This class
of structures plays the role of a statistical attractor. Moreover, its
attraction basin seems to be the whole configuration space in the
sense that, independent of the initial geometry of the system, the
dissolution-redeposition process always tends to create the same
category of geometries. The fractal DRA can be seen as a statistical
equilibrium for this process, which would maximize a ``geometrical
entropy'' still to be defined.

\bigskip
%\acknowledgments

The authors wish to acknowledge useful discussions with
B.~Mandelbrot. The Laboratoire de Physique de la Mati\`{e}re
Condens\'{e}e is a ``Unit\'{e} Mixte de Recherches du Centre National
de la Recherche Scientifique No.~7643''. This work has been partially
supported from the European Network - Fractals, under contract n$^o$
FMRX-CT98-0183.

\end{document}